\documentclass{UniVieCS_Thesis}

\usepackage{algorithm2e}

\usepackage[acronym]{glossaries}

\usepackage{listings}
\usepackage{xcolor}

\definecolor{codegreen}{rgb}{0,0.6,0}
\definecolor{codegray}{rgb}{0.5,0.5,0.5}
\definecolor{codepurple}{rgb}{0.58,0,0.82}
\definecolor{backcolour}{rgb}{0.95,0.95,0.92}

\lstdefinestyle{mystyle}{
    backgroundcolor=\color{backcolour},   
    commentstyle=\color{codegreen},
    keywordstyle=\color{magenta},
    numberstyle=\tiny\color{codegray},
    stringstyle=\color{codepurple},
    basicstyle=\ttfamily\footnotesize,
    breakatwhitespace=false,         
    breaklines=true,                 
    captionpos=b,                    
    keepspaces=true,                 
    numbers=left,                    
    numbersep=5pt,                  
    showspaces=false,                
    showstringspaces=false,
    showtabs=false,                  
    tabsize=2
}

\usepackage[multiuser,marginclue,nomargin,inline,index]{fixme}
\definecolor{ttwgreen}{RGB}{75,135,73}
\fxusetheme{colorsig}
\FXRegisterAuthor{ttw}{attw}{\color{ttwgreen}TTW} %

\newacronym{gcd}{GCD}{Greatest Common Divisor}

\newacronym{lcm}{LCM}{Least Common Multiple}

\newglossaryentry{latex}
{
    name=latex,
    description={Is a mark up language specially suited 
    for scientific documents}
}

\newglossaryentry{maths}
{
    name=mathematics,
    description={Mathematics is what mathematicians do}
}

\makeglossaries %

\usepackage[ngerman, british]{babel} %

\usepackage{pdfpages}
\usepackage{booktabs}
\usepackage{enumitem}

\usepackage{graphicx}
\usepackage{subcaption}

\usepackage{epigraph}

\usepackage{tikz}
\usepackage{pgfplots}
\pgfplotsset{compat=newest}
\usepgfplotslibrary{dateplot}
\pgfplotsset{compat=1.8}

\newcommand{\includepaper}[1]{%
  \includepdf[
    pages=-,
    scale=0.9,
    pagecommand={},
    offset=5mm 0
  ]{papers/#1}%
}

\newcommand{\descriptionsep}{%
  \vspace{0.6em}
  \hrule
  \vspace{0.6em}
}

\usepackage{filecontents}

\begin{document}

\renewcommand{\TitleSetup}{DISSERTATION / DOCTORAL THESIS}
\renewcommand{\TitleTitleSetup}{Titel der Dissertation / Title of the Doctoral Thesis\par}

\TitleTwo{\fontsize{17}{21}\selectfont Democratizing Measurement of Critical Mobile Infrastructure:\\
Security and Privacy in an Increasingly Centralized Communication Ecosystem}

\Who{Dipl.-Ing. Gabriel Karl Gegenhuber, BSc} %
\Degree{Doktor der technischen Wissenschaften (Dr. techn.)}
\Year{2026}
\ProgrammeCode{786 880}
\ProgrammeName{Informatik} %
\Supervisor{Univ.-Prof. Dipl.-Ing. Mag. Dr.techn. Edgar Weippl}
\SupervisorTwo{Univ.-Prof. DI Dr. techn. Johanna Ullrich, BSc} %
\CoSupervisor{0}

\Reviewer{Prof. Dr. Georgios Smaragdakis}
\CoReviewer{Prof. Dr. Daniel Gruß}

\Titlepage %

    \clearpage
	\frontmatter{}
    
    \thispagestyle{empty}

\setlength\epigraphwidth{0.5\textwidth}  %
\setlength\epigraphrule{0pt}             %

\renewcommand{\epigraphflush}{center}

\vspace*{\fill}

\epigraph{
\itshape
Schlage die Trommel und fürchte dich nicht,\\
Und küsse die Marketenderin!\\
Das ist die ganze Wissenschaft,\\
Das ist der Bücher tiefster Sinn.\\[1em]

Trommle die Leute aus dem Schlaf,\\
Trommle Reveille mit Jugendkraft,\\
Marschiere trommelnd immer voran,\\
Das ist die ganze Wissenschaft. %
\vspace{1em}
}{
Heinrich Heine, \emph{Doktrin}
}

\vspace*{\fill}

\clearpage

	\thispagestyle{empty}
\chapter{Acknowledgements}

This thesis was made possible by the support of many people.

I would like to express my sincere gratitude to my advisors, Edgar Weippl and Johanna Ullrich, for their continuous support throughout this endeavor and for granting me the freedom to develop and pursue my own research topics and ideas.

I am deeply grateful to Wilfried Mayer, who introduced me to the world of academia, both the good and the bad, and who was an exceptional advisor and mentor during the early stages of my PhD.
I would also like to express my sincere thanks to Adrian Dabrowski, who repeatedly enlightened me with his deep and specialized knowledge of cellular networks and taught me countless invaluable tricks in \LaTeX.

In addition, I gratefully acknowledge the Internet Stiftung for the netidee scholarship, which supported my dissertation.

Furthermore, I would like to thank my colleagues and co-authors, above all Florian, but also Markus and David, who accompanied me throughout my PhD journey.
Working with you was not only productive but always great fun.
I would also like to thank Michael Pucher, who supported me on numerous teaching assignments and consistently placed the needs of his students before his own.

Philipp Frenzel was a fantastic collaborator and student assistant.
Over the years, he contributed significantly to advancing MobileAtlas and to many of the publications included in this dissertation.
I am excited to see all the things he will achieve in the coming years.
Together with Maximilian Günther, we spent many late but very enjoyable nights in the lab researching vulnerabilities, creating measurements, and eating pickles (or pizza at Pagliacci's).

I also greatly enjoyed working with Aljosha Judmayer, who always knew how to keep our research within ethical boundaries, thereby helping us avoid legal trouble, and who was, in addition, the ideal person for discussions on cryptography.

I would also like to thank Barbara Limbeck-Lilienau, for juggling an impressive workload while always taking the time to provide clear and thorough answers to administrative questions.

I am deeply grateful to my partner Nadine Caroline for reminding me that there is a life beyond a computer screen and for ensuring that I take time to rest and spend time in nature.
Finally, I would like to thank my parents, Eva and Karl, for awakening and encouraging my curiosity at a young age and my sisters, Julia and Hannah, for supporting me throughout my life.

\clearpage
	\chapter{Abstract}
Cellular networks serve as the backbone of global communication, providing critical access to telephony and the Internet, often in regions lacking alternatives.
However, the growing complexity of these networks, driven by architectural innovations (e.g., Voice over IP, eSIMs) and commercial dynamics (e.g., roaming, virtual operators, zero-rating), remains poorly understood due to the lack of open, scalable, and geographically diverse measurement tools and independent measurement studies.

Moreover, access to mobile networks today is no longer limited to the traditional radio interface.
Technologies like Voice-over-WiFi (VoWiFi) offer alternative connectivity paths via third-party Internet infrastructure, extending operator reach into environments with limited cellular coverage.
At the same time, over-the-top (OTT) messaging services such as WhatsApp and Signal have become central to modern communication, accounting for a substantial share of global messaging and voice traffic while bypassing traditional operator-controlled channels entirely.

This dissertation addresses these challenges by introducing new approaches for independent, scalable, and reproducible measurements of mobile communication systems without requiring cooperation from network or platform operators.
We design, implement, and open-source measurement platforms that enable controlled experiments across cellular radio networks, operator-provided services, and OTT messaging applications.
Using these tools, we conduct multi-layer empirical studies and uncover security- and privacy-relevant weaknesses, including inconsistencies in roaming billing and traffic classification, insecure VoWiFi configurations, and metadata leaks in widely used messaging platforms that enable silent user monitoring and denial-of-service attacks.
Overall, this dissertation demonstrates that independent, active measurements are essential for understanding the evolving cellular communication system.
It provides practical tools and empirical evidence that increase transparency and support future research into the security and privacy of modern mobile communication systems.

	\chapter{Kurzfassung}
Mobilfunknetze bilden das Rückgrat der globalen Kommunikation und ermöglichen den Zugang zu Telefondiensten und dem Internet, häufig auch in Regionen, in denen keine alternativen Zugangstechnologien verfügbar sind.
Die zunehmende Komplexität dieser Netze, bedingt durch architektonische Innovationen (z. B. Voice over IP, eSIMs) sowie durch kommerzielle Dynamiken (z. B. Roaming, virtuelle Netzbetreiber, Zero-Rating), ist jedoch bislang nur unzureichend erforscht.
Ein wesentlicher Grund hierfür ist das Fehlen offener, skalierbarer und geografisch diverser Messwerkzeuge sowie unabhängiger empirischer Messstudien.

Darüber hinaus ist der Zugang zu Mobilfunkdiensten heute nicht mehr ausschließlich auf die klassische Funkschnittstelle beschränkt.
Technologien wie Voice-over-WiFi (VoWiFi) ermöglichen alternative Zugangspfade über das Internet und erweitern damit die Reichweite von Netzbetreibern auf Umgebungen mit eingeschränkter Netzabdeckung. Gleichzeitig haben sich sogenannte Over-the-Top-(OTT-)Messaging-Dienste wie WhatsApp und Signal zu zentralen Bestandteilen moderner Kommunikation entwickelt.
Sie tragen einen erheblichen Anteil des weltweiten Nachrichten- und Sprachverkehrs, ohne dabei auf die traditionellen, von Mobilfunkbetreibern kontrollierten Signalisierungs- und Abrechnungsstrukturen zurückzugreifen.

Diese Dissertation adressiert die genannten Herausforderungen durch die Entwicklung neuer Ansätze für unabhängige, skalierbare und reproduzierbare Messungen mobiler Kommunikationssysteme, ohne eine Kooperation mit Netz- oder Plattformbetreibern vorauszusetzen.
Hierzu werden Messplattformen entworfen, implementiert und als Open-Source-Software veröffentlicht, die kontrollierte Experimente über Mobilfunkzugangsnetze, betreiberseitige Dienste sowie OTT-Messaging-Anwendungen hinweg ermöglichen.
Mithilfe dieser Werkzeuge werden umfassende empirische Studien durchgeführt, die sicherheits- und privatsphärerelevante Schwachstellen aufdecken, darunter Inkonsistenzen bei Roaming-Abrechnung und Traffic-Klassifikation, unsichere VoWiFi-Konfigurationen, sowie Metadatenleaks in weit verbreiteten Messaging-Plattformen, die eine unbemerkte Überwachung von Nutzer:innen und Denial-of-Service-Angriffe ermöglichen.

Insgesamt zeigt diese Arbeit, dass unabhängige, aktive Messungen unerlässlich sind, um das sich wandelnde Mobilfunkkommunikationssystem zu verstehen. Sie stellt praxisnahe Werkzeuge und empirische Erkenntnisse bereit, die die Transparenz erhöhen und zukünftige Forschung zur Sicherheit und zum Schutz der Privatsphäre moderner mobiler Kommunikationssysteme unterstützen.

\clearpage
	\pagebreak
	
    \microtypesetup{protrusion=false}
    \tableofcontents{}
    \cleardoublepage{}
    \microtypesetup{protrusion=true}
    
	\pagebreak
	
	\mainmatter{}
    \chapter{Introduction}
\label{chap:1} 
Cellular networks are a primary access technology for both the public telephone system and the Internet --- often the only available option in many parts of the world~\cite{gsma_2024_state}.
Beyond their role in personal and business communication, they are vital for crisis response and emergency scenarios.
Technologies such as roaming, virtual network operators (MVNOs), and travel (e)SIMs interconnect previously isolated infrastructures, forming a compound system embedded in a complex global web. 
Free-roaming agreements --- most notably within the European Union --- further blur national boundaries, offering users a seamless and unified mobile experience across borders.

Despite their critical role in today's society, cellular networks remain among the least transparent and least independently measurable components of the global Internet.
This limited measurability is not a consequence of limited adoption, but of the absence of controlled and scalable measurement tools that reflect the unique architecture and global scope of cellular access networks.
As a result, independent insight into provider practices --- particularly under roaming conditions --- remains severely limited.

In addition to traditional 3GPP-specified telephony and messaging, third-party instant messaging platforms such as WhatsApp play a central role in today's mobile communication ecosystem.
Although they carry a substantial share of message traffic, these services are developed and controlled by private companies, introducing similar transparency and auditability challenges at the application layer, and ultimately requiring a significant degree of blind trust.

To improve transparency, security, and resilience in this sector, independent and accessible measurement tools are urgently needed --- tools that can serve researchers, regulators, and industry stakeholders alike for auditing and controlled testing.
As complexity increases, so do potential vulnerabilities and corner cases.
Without proper tools to introspect and analyze the inner workings of these networks and applications, critical issues may remain undetected, compromising both security and infrastructure resilience.

To address this gap, this dissertation proposes new measurement tools and presents independent empirical studies of mobile communication systems, examining multiple layers of the ecosystem, from the radio access layer to VoWiFi and third-party instant messaging services, with the common goal of enabling independent and reproducible measurements and strengthening security and privacy.

\section{Thesis Statement}
\noindent\textit{This dissertation argues that democratized, independent measurement (i.e., controlled and scalable experiments without operator cooperation) enables the systematic discovery and quantification of security, privacy, and policy failures across mobile communication layers---failures that have direct consequences for end users and would otherwise remain invisible.}

\medskip
Across the presented case studies, the key methodological contribution is not merely measurement at scale, but \emph{independent} measurement: experiments that remain feasible even when operators or platforms do not collaborate, and that can be replicated across countries, providers, and time.
Concretely, this dissertation argues that democratized measurement enables:
\begin{itemize}
    \item \textbf{coverage} across operators, countries, and deployment settings,
    \item \textbf{control} over experimental conditions without privileged operator access or platform cooperation, and
    \item \textbf{comparability} of results across technology-defined layers through standardized, repeatable experiments and shared artifacts.
\end{itemize}
These properties enable systematic discovery and quantification of vulnerabilities and risks that are unlikely to surface through documentation or the operators and platforms themselves.

\section{Cellular Network Research Gap}
To evaluate the thesis statement across layers, this dissertation combines reusable measurement artifacts with empirical studies.
Each layer, defined by its underlying technology, implies different trade-offs in independence, control, and scale and exposes distinct security and privacy risks.

For measurements in the cellular network domain, we distinguish between passive and active measurements.

Measurements with \textbf{passive data} are often conducted with the help of industry partners, e.g., Internet Service Providers (ISPs) or Mobile Network Operators (MNOs).
These measurements rely on data collected from operational networks, typically at a large scale.
However, such data is rarely publicly accessible and generally requires privileged access to core infrastructure. As a result, passive measurements are largely inaccessible to independent researchers or early-stage academic efforts.

On the other hand, \textbf{active measurements} have proven to be a vital tool for conducting independent and open research.
The two most common practices for active measurements include (i) in-situ- or in-vivo measurements, and (ii) exclusive measurement setups.

\paragraph{App-based approach.}
\textit{In-situ} (``at location'') and \textit{in-vivo} (``within the system'') measurements often run on a (volunteer's) phone that is also used for other tasks.
This method increases coverage by reducing the financial burden of participating in a given study, thereby adding more measurement units.
While appropriate for, e.g., user studies, this method might impact the accuracy of technical measurements since the non-exclusivity blurs the distinction between measurement (\textit{signal}) and background- or user activity (\textit{noise}).
The user's mobility also interferes with the common goal of steady (consistent and repeatable) measurement environments. 
Volunteers also bear responsibility for any billing charges incurred during the experiment, rendering this approach impractical for certain scenarios, particularly those involving roaming or phone calls.
Moreover, the measurement operator often lacks full control over key parameters, such as the access technology in use (e.g., 3G, 4G, or 5G), and has limited visibility into critical data.
For instance, billing statements are typically accessible only to the volunteer, restricting opportunities for external validation and comprehensive analysis.

\paragraph{Deploying dedicated hardware.}
Exclusive measurement setups require a separate test unit in a controlled environment.
While some external factors (such as the network's utilization or radio noise) are commonly outside of the researchers' sphere of influence, other factors can be controlled; this includes the distance to a cell tower, the used access technology, and background data usage.

While deploying, operating, and maintaining a large fleet of distributed measurement probes remains a logistical challenge, such platforms can, in principle, be democratized and scaled through community-driven efforts.
This model has proven successful in the context of independent fixed-line Internet measurements, most notably through community-driven efforts like RIPE Atlas~\cite{staff_2015_ripe, anderson_2014_global}. %

Replicating this model in the cellular domain, however, presents fundamental challenges.
Requiring the probe host to supply and manage their own SIM card introduces billing liability, which not only deters participation but also severely limits the number of network operators measurable from a single probe.
Supporting a diverse set of SIM cards across different locations or within roaming scenarios does not scale, thereby impeding broader deployment and coverage.

Taken together, these constraints create a structural barrier for independent research, particularly for academic groups, regulators, and researchers without privileged operator access.
Meaningful security and privacy analysis of cellular networks requires controlled, repeatable, and globally diverse measurements, precisely the properties that existing approaches fail to provide.
Without these properties, security flaws, billing inconsistencies, and privacy leaks may remain systematically invisible.

In summary, existing measurement approaches force researchers to trade off scalability, control, and independence.
Consequently, none of these approaches is sufficient for conducting controlled, repeatable, and globally diverse measurements of cellular networks without operator cooperation.

\section{Research Questions}
\label{sec:research-questions}
Motivated by the limitations of existing measurement approaches and guided by the thesis statement above, this dissertation investigates the following research questions:

\begin{enumerate}
    \item[RQ1] How can cellular networks be measured independently, in a controlled manner, and at scale, without requiring cooperation from network operators?
    \item[RQ2] How do roaming scenarios, alternative access technologies, and operator billing mechanisms interact in real-world cellular network deployments?
    \item[RQ3] How do security and privacy properties, as well as measurement capabilities, differ across layers of the mobile communication ecosystem, ranging from cellular access networks to operator-provided services and OTT messaging platforms?
\end{enumerate}

\medskip

RQ1 and RQ2 focus on how independent measurement can characterize operator behavior and operator-provided services without requiring cooperation.
RQ3 applies the same measurement perspective to increasingly centralized OTT messaging platforms, where Internet-scale observation can reveal ecosystem-wide attack surfaces and privacy risks.

RQ1 is addressed by Chapters~\ref{chap:2}--\ref{chap:3} (measurement platforms and scalability).\\
RQ2 is addressed by Chapters~\ref{chap:3}--\ref{chap:6} (roaming, billing, alternative access).\\
RQ3 is addressed by Chapters~\ref{chap:5}--\ref{chap:8} (cross-layer security/privacy properties and observability).
Chapter~\ref{chap:4} departs from cellular measurements and demonstrates how a distributed community-driven measurement platform (RIPE Atlas) can reveal Internet centralization that affects anonymity in Tor, reinforcing the dissertation's broader theme of democratized, independent measurements and the risks introduced by centralization in global communication systems.

\section{New Measurement Approaches and Results}
To answer these research questions, this dissertation introduces new approaches to large-scale (i.e., international) security and privacy measurements in cellular networks and mobile communication systems.
It follows a multi-layer measurement perspective, where each layer targets a different part of the mobile communication ecosystem and exposes distinct measurement challenges and security properties.
This progression from radio access networks to VoWiFi and ultimately to OTT messaging platforms mirrors the ongoing shift of control and observability from decentralized, operator-bound infrastructure toward globally centralized, Internet-based platforms.

\begin{figure}[bth]
    \centering
    \begin{minipage}[t]{0.32\linewidth}
        \centering
        \textbf{Angle 1}\\[0.8em]
        \includegraphics[height=0.35\linewidth]{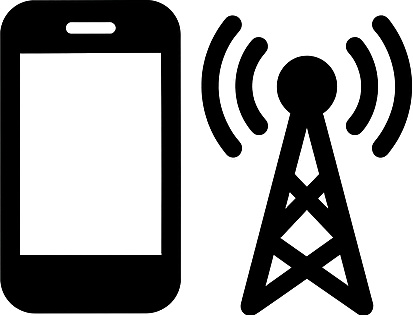}\\[0.8em]
        \textbf{Radio Layer}\\[0.4em]
        {\small Access via cellular network\\
        Backend by operator}
    \end{minipage}
    \hfill
    \begin{minipage}[t]{0.32\linewidth}
        \centering
        \textbf{Angle 2}\\[0.8em]
        \includegraphics[height=0.35\linewidth]{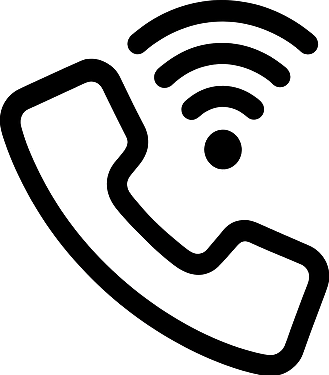}\\[0.8em]
        \textbf{VoWiFi Layer}\\[0.4em]
        {\small Access via Wi-Fi AP\\
        Backend by operator}
    \end{minipage}
    \hfill
    \begin{minipage}[t]{0.32\linewidth}
        \centering
        \textbf{Angle 3}\\[0.8em]
        \includegraphics[height=0.35\linewidth]{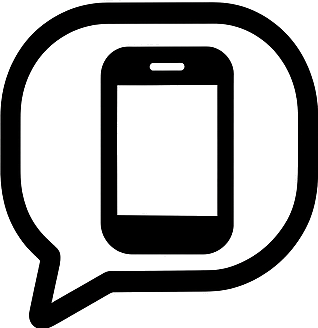}\\[0.8em]
        \textbf{Instant Messengers}\\[0.4em]
        {\small Access via Wi-Fi AP\\
        Backend by third party}
    \end{minipage}

    \caption{Overview of the three research angles addressed in this work. Large-scale measurements at the radio access layer require a local point of presence for connectivity to cellular base stations, which imposes significant logistical and deployment challenges, while Wi-Fi calling and instant messaging rely on standard Internet connections.
    Although the first two technologies are standardized by 3GPP, all three lack adequate tooling for comprehensive auditing and independent security testing.
    The progression from radio access networks to VoWiFi and finally to OTT messaging platforms reflects an increasing degree of centralization, in which control and observability shift from decentralized, operator-bound infrastructure toward globally centralized, Internet-based services.
    }
    
    \label{fig:research-topics}
\end{figure}

Figure~\ref{fig:research-topics} illustrates the three complementary research perspectives considered in this work: the cellular radio access layer, the Voice over Wi-Fi (VoWiFi) layer, and third-party instant messaging services.
Each angle exposes a different form of centralization, opacity, or control.

\subsection{Democratizing Radio Layer Measurements}
Large-scale and controlled experiments at the cellular radio access layer require a local point of presence to interact with base stations, making such measurements difficult to scale across operators, countries, and roaming scenarios.
Existing approaches typically rely on physically co-locating subscriber identities (i.e., SIM cards) with measurement devices, which introduces significant logistical and operational constraints.
As a result, repeatable radio-layer measurements remain largely inaccessible to independent researchers.

\begin{figure}[bt]
  \centering
  \includegraphics[width=0.65\linewidth]{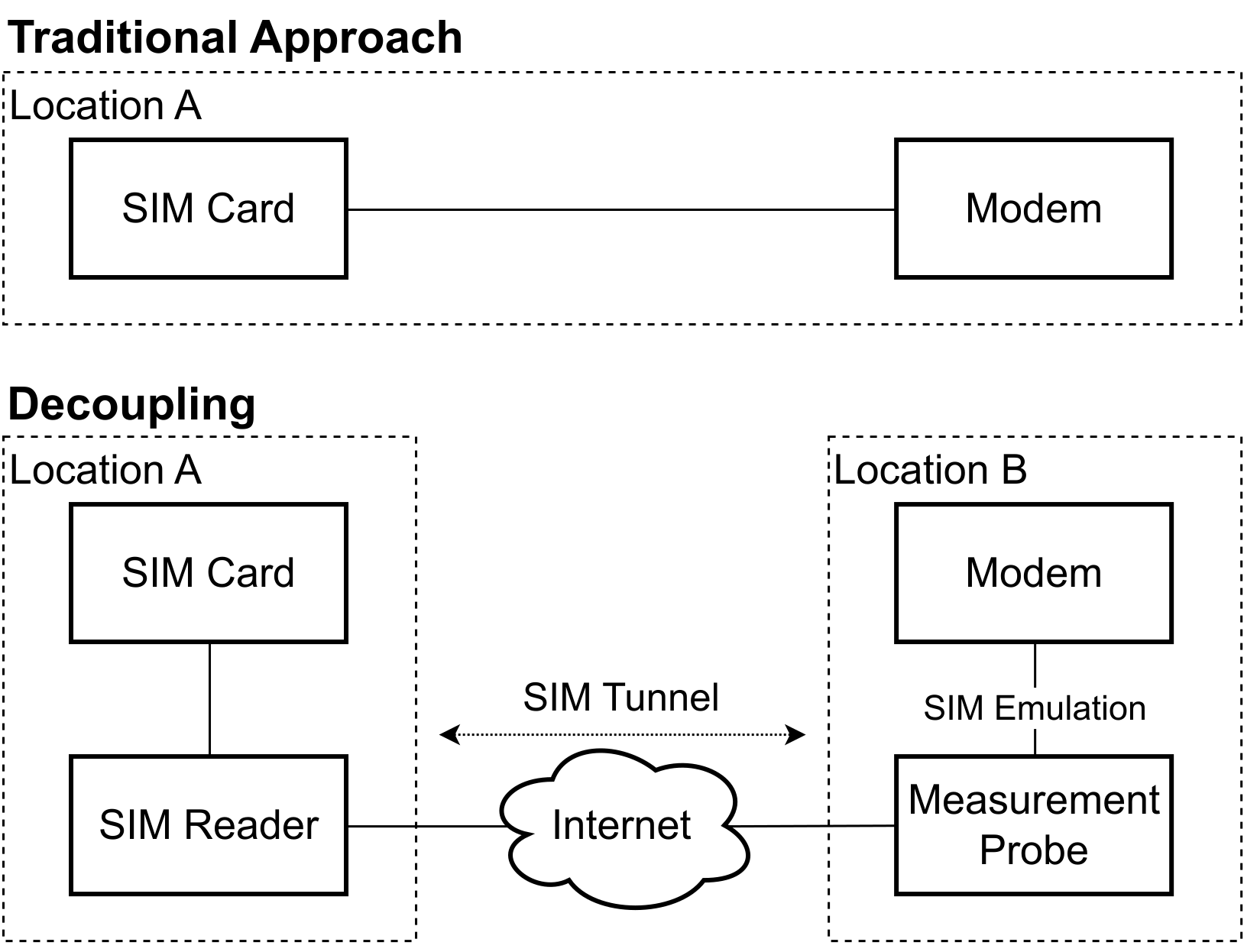}
  \caption{SIM card and modem are geographically decoupled via a SIM tunnel relaying the protocol over the Internet.}
  \label{fig:sim-tunnel}
\end{figure}

To address this gap in dedicated measurement platforms for cellular networks, this dissertation introduces the \textsc{MobileAtlas} framework~\cite{gegenhuber_2023_mobileatlas}, based on low-cost hardware (e.g., a Raspberry~Pi~4 and a COTS cellular modem).
Conceptually, \textsc{MobileAtlas} is inspired by RIPE Atlas: both aim to democratize measurement by enabling distributed experiments from many independent vantage points.
As shown in Figure~\ref{fig:sim-tunnel}, \textsc{MobileAtlas} geographically decouples the SIM card from the modem by tunneling the SIM card's protocol over the Internet and emulating its signal at the modem.
With this approach, measurement probes and SIM cards can be at different locations and thus flexibly shared with other participants, without the need for permanently installed SIM cards or any physical movement of components.
A SIM card hosted at a fixed location can be virtually connected to measurement probes around the world within seconds, each offering a local breakout to its respective cellular infrastructure.
This enables the SIM to be quickly evaluated within diverse network environments.
Additionally, the platform utilizes Linux namespaces to isolate the modem connection from any background activity and provide sterile measurement access for fine-grained measurements.

While SIM tunneling inherently introduces additional round-trip latency, authentication and session key generation remain robust to such delays, as they are designed to tolerate network congestion and retransmissions caused by weak signal conditions.
We validate this robustness experimentally across a wide range of operators, including intercontinental tunneling scenarios.

Furthermore, since all ISO 7816~\cite{iso7816-3} APDU commands between the modem and SIM card are relayed through the tunnel, the corresponding traffic can be inspected to gain further insights into protocol behavior and SIM-card interactions (e.g., proactive SIM commands).
The resulting measurement platform provides a controlled environment, scalability, and cost-effectiveness.
Moreover, it is extensible and fully open-source\footnote{\url{https://github.com/sbaresearch/mobile-atlas}}, enabling other researchers to contribute locations, SIM cards, and measurement scripts.

In an initial study~\cite{gegenhuber_2023_mobileatlas}, we use the platform to inspect and compare network configurations in both domestic and (home-routed) roaming scenarios.
This analysis revealed weak firewall configurations during operator migrations from Carrier-Grade NAT (CGNAT) to IPv6-based architectures.
Additionally, we uncovered instances of hidden SIM card activity, such as silent SMS messages to the operator sent via proactive SIM commands.
To further investigate signaling behavior, we generated test calls and analyzed subtle variations in ringback tones.
These differences enable detailed fingerprinting of mobile operators and can inadvertently reveal a user's operator and, by extension, their country-level location.

Moreover, we use the platform for fine-grained traffic experiments, investigating operators' billing mechanisms under differential pricing schemes (i.e., zero-rating offers)~\cite{gegenhuber_2022_zero}.
Our analysis reveals potentially problematic behavior (i.e., inadvertent billing of zero-rated traffic) at nearly all operators examined and identifies possible vectors for free-riding attacks (e.g., spoofed Host- or SNI- headers).

\paragraph{Low-cost SIM tracing.}
Since the SIM tunnel relies on low-cost, readily available peripherals (i.e., a UART interface and GPIO ports), we further reduce the economic barrier by implementing and publishing the design for the Raspberry~Pi~Pico~\cite{gegenhuber_2025_simulator}.
This enables SIM-tracing capabilities (i.e., inspecting, rewriting, and relaying traffic) at a hardware cost of as little as 5\,USD.
In addition to supporting physical SIM cards, our system also supports \textit{in vitro} analysis of eSIMs by leveraging the SIM Access Profile (SAP)~\cite{bluetooth_sim_access_2003} available on Android devices.

\subsection{Leveraging Alternative Access Technologies}
\begin{figure}[bt]
  \centering
  \includegraphics[width=0.8\linewidth]{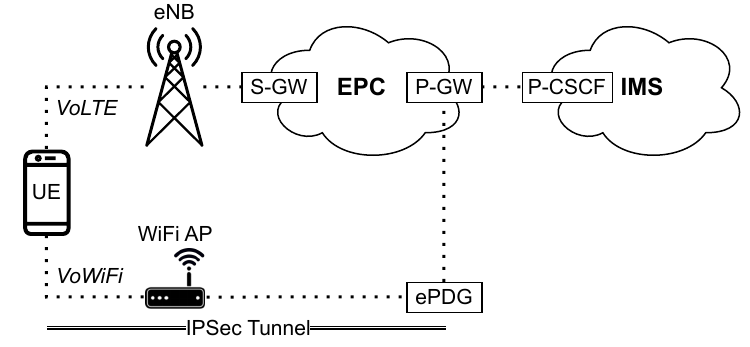}
  \caption{(Simplified) LTE network architecture for VoLTE and VoWiFi. The ePDG server is exposed to the Internet.}
  \label{fig:architecture-volte-vowifi}
\end{figure}

In current cellular network generations (4G, 5G) the IMS (IP Multimedia Subsystem) plays an integral role in terminating voice calls and short messages.
Many operators use VoWiFi (Voice over Wi-Fi, also Wi-Fi calling) as an alternative network access technology to complement their cellular coverage in areas where no radio signal is available (e.g., rural territories or shielded buildings).
In practice, VoWiFi is often prioritized over VoLTE, making it a primary calling path in many deployments and increasing its relevance for communication security and privacy.
Since this technology requires operators to publicly expose parts of their infrastructure to the Internet, it also creates a new measurement vector for large-scale cellular measurements (cf. ePDG in Figure~\ref{fig:architecture-volte-vowifi}).
We leverage these alternative access technologies by conducting Internet-based scanning and reconnaissance measurements, thereby offering a comprehensive global perspective on current deployment states and operational practices.

Our work evaluates the current deployment status of VoWiFi among worldwide operators and uses commercial VPN subscriptions to analyze existing geoblocking measures on the IP layer~\cite{gegenhuber_2024_geoblocking}.
More specifically, we show that a substantial share of operators implement geoblocking at the DNS- or VoWiFi protocol level, and highlight severe drawbacks in terms of emergency calling service availability.

Moreover, we found critical vulnerabilities in commercial VoWiFi implementations (e.g., security downgrade attacks~\cite{gegenhuber_2024_diffie}) and configurations (e.g., outdated ciphers~\cite{gegenhuber_2024_never}, private-key reusage~\cite{gegenhuber_2024_diffie}), both affecting the communication security and privacy of hundreds of millions of users.

We open-source both our VPN-driven scanning solution for rapid distributed Internet measurements\footnote{\url{https://github.com/sbaresearch/scanywhere/}} and our scripts to evaluate global VoWiFi deployments and configurations\footnote{\url{https://github.com/sbaresearch/vowifi-epdg-scanning}}.

\subsection{Evaluating Over-the-Top Messaging Applications}
Beyond the radio layer and operator-controlled services such as VoWiFi or RCS, a significant share of voice and messaging traffic is now carried by OTT messaging platforms.
Popular applications like WhatsApp, iMessage, and Signal account for a large portion of global communication volume, bypassing traditional signaling infrastructures and operator billing mechanisms entirely.
Evaluating the security, privacy, and network behavior of these platforms is essential, as they shape user experience and often serve as the primary channel in regions with censorship, surveillance, or restricted network access.
However, since many of these applications are closed-source and developed by commercial entities, comprehensive auditing and in-depth evaluation of their security and privacy properties remain challenging.
Although OTT platforms are not cellular technologies, they now dominate user-facing mobile communication and must therefore be included in any realistic and comprehensive security analysis.

Despite offering end-to-end encryption (E2EE), our measurements show that many widely used messaging applications remain susceptible to privacy leaks.
For example, refilling mechanisms for ephemeral encryption keys can unintentionally expose a user's online status on individual devices~\cite{gegenhuber_2025_prekeypogo} and reveal information about their used operating system.
In addition, global synchronization issues in backend infrastructures can lead to denial-of-service (DoS) conditions for targeted users.
To support independent verification and raise user awareness, we open-source a research prototype that enables querying key bundles for arbitrary numbers registered on WhatsApp\footnote{\url{https://github.com/sbaresearch/prekey-pogo}}.

Moreover, acknowledgment messages for successful message transmission (i.e., delivery receipts) can be used to silently, precisely, and consistently monitor a victim's connection round-trip time (RTT), thereby leaking activity states or behavioral patterns that may disclose a person's location or routines~\cite{gegenhuber_2025_carelesswhisper}.
Such \textit{silent pings} can also be abused for resource exhaustion attacks, allowing adversaries to drain a user's battery or consume mobile data quota without their awareness.

Recent work shows that dominant OTT messaging platforms (such as WhatsApp) may increase centralization and give rise to enumeration vulnerabilities, exposing roughly 3.5 billion phone numbers and corresponding to a significant fraction of the global mobile user base~\cite{gegenhuber_2025_heythere}.

These findings highlight the close interplay between OTT applications and cellular networks and underline the need to examine both layers as part of a comprehensive security and privacy analysis.

\section{Contributions}
This dissertation advances the state of independent cellular network measurement by developing new measurement approaches and providing empirical insights into modern mobile communication systems across multiple layers.
In line with the thesis statement, it shows that democratized, controlled, and scalable measurements can reveal and quantify cross-layer security, privacy, and policy failures without operator cooperation.
Collectively, these contributions establish a general methodology for independent, scalable security measurements in mobile communication systems.
The main contributions of this work are as follows:

\begin{itemize}
  \item \textbf{A scalable and independent platform for radio-layer cellular measurements.}
  This dissertation introduces \textsc{MobileAtlas}, a low-cost, open-source measurement framework that enables controlled and repeatable cellular measurements without operator cooperation by geographically decoupling SIM cards from measurement probes.

  \item \textbf{Empirical analysis of roaming behavior and billing mechanisms.}
  Using \textsc{MobileAtlas}, this work provides fine-grained measurements of domestic and roaming scenarios, uncovering inconsistencies in IP addressing, firewall configurations, proactive SIM behavior, and traffic classification under zero-rating schemes.

  \item \textbf{Large-scale evaluation of VoWiFi availability, geoblocking, and security.}
  This work presents a global-scale analysis of commercial VoWiFi deployments, revealing deployment gaps, geoblocking practices, and security-relevant misconfigurations affecting service availability, emergency calling, and user privacy for subscribers worldwide.

  \item \textbf{Systematic exposure of security and privacy weaknesses in prevalent OTT messaging applications.}
  This dissertation investigates widely used instant messaging applications, showing how protocol design and backend behavior can leak metadata, enable silent monitoring, and expose users to denial-of-service and enumeration risks.
  
  \item \textbf{Open-source tools and datasets for reproducible mobile network research.}
  Measurement platforms and analysis tools developed in this dissertation are released as open source, enabling independent validation, reuse, and extension by the research community.
\end{itemize}

\section{Publications (Conference and Journal Papers)}
\begin{description}[leftmargin=1cm, style=nextline]
\item[\cite{gegenhuber_2023_mobileatlas}] 
\underline{Gabriel K. Gegenhuber}, Wilfried Mayer, Edgar Weippl, Adrian Dabrowski.  
\textit{MobileAtlas: Geographically Decoupled Measurements in Cellular Networks for Security and Privacy Research.}  
In 32nd USENIX Security Symposium (USENIX Security), 2023.

\item[\cite{gegenhuber_2022_zero}] 
\underline{Gabriel K. Gegenhuber}, Wilfried Mayer, Edgar Weippl.  
\textit{Zero-Rating, One Big Mess: Analyzing Differential Pricing Practices of European MNOs.}  
In IEEE Global Communications Conference (GLOBECOM), 2022.

\item[\cite{gegenhuber_2023_tor}] 
\underline{Gabriel K. Gegenhuber}, Markus Maier, Florian Holzbauer, Wilfried Mayer, Georg Merzdovnik, Edgar Weippl, Johanna Ullrich.  
\textit{An Extended View on Measuring Tor AS-level Adversaries.}  
In Computers \& Security (COSE) 132, 2023.

\item[\cite{gegenhuber_2024_geoblocking}] 
\underline{Gabriel K. Gegenhuber}, Philipp É. Frenzel, Edgar Weippl.  
\textit{Why E.T. Can't Phone Home: A Global View on IP-based Geoblocking at VoWiFi.}  
In 22nd Annual International Conference on Mobile Systems, Applications and Services (MobiSys), 2024.

\item[\cite{gegenhuber_2024_diffie}] 
\underline{Gabriel K. Gegenhuber}, Florian Holzbauer, Philipp Frenzel, Edgar Weippl, Adrian Dabrowski.  
\textit{Diffie-Hellman Picture Show: Key Exchange Stories from Commercial VoWiFi Deployments.}  
In 33rd USENIX Security Symposium (USENIX Security), 2024.

\item[\cite{gegenhuber_2025_carelesswhisper}] 
\underline{Gabriel K. Gegenhuber}, Maximilian Günther, Markus Maier, Aljosha Judmayer, Florian Holzbauer, Philipp É. Frenzel, Johanna Ullrich.  
\textit{Careless Whisper: Exploiting Silent Delivery Receipts to Monitor Users on Mobile Instant Messengers.}  
In 28th International Symposium on Research in Attacks, Intrusions and Defenses (RAID), 2025.  
\textbf{Distinguished with the Best Paper Award.}

\item[\cite{gegenhuber_2025_prekeypogo}] 
\underline{Gabriel K. Gegenhuber}, Philipp É. Frenzel, Maximilian Günther, Aljosha Judmayer.  
\textit{Prekey Pogo: Investigating Security and Privacy Issues in WhatsApp's Handshake Mechanism.}  
In 19th USENIX WOOT Conference on Offensive Technologies (WOOT), 2025.

\descriptionsep

\item[\cite{gegenhuber_2025_heythere}] 
\underline{Gabriel K. Gegenhuber}, Philipp É. Frenzel, Maximilian Günther, Johanna Ullrich, Aljosha Judmayer.  
\textit{Hey there! You are using WhatsApp: Enumerating Three Billion Accounts for Security and Privacy.}  
In 33rd Annual Network and Distributed System Security Symposium (NDSS), 2026.
\textbf{Complementary work, not part of the thesis.}

\end{description}

\section{Workshops, Extended Abstracts, and Posters}

\begin{description}[leftmargin=1cm, style=nextline]
\item[\cite{gegenhuber_2024_never}] 
\underline{Gabriel K. Gegenhuber}, Philipp É. Frenzel, Edgar Weippl.  
\textit{Never Gonna Give You Up: Exploring Deprecated NULL Ciphers in Commercial VoWiFi Deployments.}
Poster \& Extended Abstract.
In 17th ACM Conference on Security and Privacy in Wireless and Mobile Networks (WiSec), 2024.

\item[\cite{gegenhuber_2025_scanywhere}] 
\underline{Gabriel K. Gegenhuber}, Philipp É. Frenzel.  
\textit{Scanywhere: Distributed Internet Scanning Leveraging Commercial VPN Subscriptions.}
Poster \& Extended Abstract.
In 9th Network Traffic Measurement and Analysis Conference (TMA), 2025.
\textbf{Distinguished with the Best Poster Award.}

\item[\cite{gegenhuber_2025_simulator}] 
\underline{Gabriel K. Gegenhuber}, Philipp É. Frenzel, Adrian Dabrowski.
\textit{SIMulator: SIM Tracing on a (Pico-)Budget.}  
Poster \& Extended Abstract.
In 18th ACM Conference on Security and Privacy in Wireless and Mobile Networks (WiSec), 2025.

\item[\cite{gegenhuber_2025_novel}] 
\underline{Gabriel K. Gegenhuber}.
\textit{Security and Privacy Measurements in Cellular Networks: Novel Approaches in a Global Roaming Context.}  
Extended Abstract (Doctoral Symposium).
In 32nd ACM Conference on Computer and Communications Security (CCS), 2025.

\item[\cite{gegenhuber_2025_airtag}] 
\underline{Gabriel K. Gegenhuber}, Leonid Liadveikin, Florian Holzbauer, Sebastian Strobl.
\textit{A Relay a Day Keeps the AirTag Away: Practical Relay Attacks on Apple's AirTags.}  
Poster \& Extended Abstract.
In 41st Annual Computer Security Applications Conference (ACSAC), 2025.
\end{description}

\section{Artifacts and Impact}
\label{sec:artifacts}

This dissertation follows an artifact-driven research approach, where measurement studies are accompanied by open-source tools and responsible disclosure procedures.

\subsection{Open Source Projects and Artifacts}

\begin{description}
    \item \href{https://github.com/sbaresearch/mobile-atlas}{\textbf{mobile-atlas}}: Cellular measurement platform for scalable roaming measurements
    \item \href{https://github.com/sbaresearch/scanywhere}{\textbf{scanywhere}}: Global Internet scanning solution driven by commercial VPN solutions
    \item \href{https://github.com/sbaresearch/vowifi-epdg-scanning}{\textbf{vowifi-epdg-scanning}}: Evaluating VoWiFi cipher and key exchange methods in commercial networks
    \item \href{https://github.com/sbaresearch/mbn-mcfg-tools}{\textbf{mbn-mcfg-tools}}: Parsing and packing proprietary Qualcomm MBN files (used for modem configurations)
    \item \href{https://github.com/sbaresearch/prekey-pogo}{\textbf{prekey-pogo}}: Requesting device directory and prekey information of arbitrary WhatsApp users
    \item \href{https://github.com/gommzystudio/device-activity-tracker}{\textbf{device-activity-tracker}}: Monitoring WhatsApp devices via silent pings. \emph{This project is based on our publication, but was developed by an external party.}
\end{description}

\subsection{Impact and Responsible Disclosure}
Prior to the public release of the publications and artifacts included in this dissertation, all identified security and privacy issues were reported to the affected parties in accordance with responsible disclosure practices. 

The findings of this work were subsequently shared with the broader security research and practitioner community through presentations at two DEF CON conferences, one Black Hat conference, and several additional venues (e.g., in front of the GSMA Fraud and Security Group).

In addition, the disclosed vulnerabilities and their implications were reported by high-profile international media outlets, contributing to broader public awareness of security and privacy risks in mobile communication systems.

The following vulnerability identifiers and disclosures are directly associated with the results presented in this dissertation:

\begin{description}
    \item \textbf{CVE-2025-20647 (medium severity)}: MediaTek, VoLTE/VoWiFi Denial of Service due to NULL Pointer Dereference at SIP
    \item \textbf{CVE-2024-22064 (high severity)}: ZTE, Private-Key Sharing among Global Operators
    \item \textbf{CVE-2024-20069 (high severity)}: MediaTek, Downgrade Vulnerability at VoWiFi IKE Protocol
    \item \textbf{CVD-2024-0089}: GSMA, Deprecated VoWiFi Configurations and Coordinated Disclosure of CVE-2024-22064, CVE-2024-20069
    \item \textbf{Signal}: The official Signal GitHub repository contains an \href{https://github.com/signalapp/Signal-Android/pull/14463}{ongoing discussion} regarding potential countermeasures against silent ping attacks.
    \item \textbf{WhatsApp}: Security reports submitted as part of this work were acknowledged by WhatsApp and are currently in the process of being addressed.
\end{description}

\section{How to Read This Dissertation}
\label{sec:readers-guide}
This dissertation is a cumulative thesis. Chapters~\ref{chap:2}--\ref{chap:8} correspond to peer-reviewed publications that are included verbatim to preserve their original technical context and reproducibility.
Readers primarily interested in the overarching narrative can focus on Chapter~\ref{chap:1} (motivation, thesis statement, research questions, and contributions) and Chapter~\ref{chap:9} (synthesis and outlook), and consult Chapters~~\ref{chap:2}--\ref{chap:8} selectively for technical details and empirical evidence.

\paragraph{A multi-layer measurement perspective.}
This dissertation adopts a layered view of the mobile communication ecosystem.
Across these layers, the central methodological goal is to enable \emph{independent, controlled, and scalable} measurements without requiring cooperation from operators or platform providers.
This progression reflects a shift in where control and observability reside: from decentralized, operator-bound infrastructure at the radio access layer, to operator-managed services exposed via the public Internet (e.g., VoWiFi), and finally to globally centralized OTT messaging platforms.

\paragraph{Chapter dependencies and reading order.}
Chapters~\ref{chap:2} and~\ref{chap:3} introduce and operationalize the dissertation's core measurement capability at the radio layer, enabling repeatable experiments across countries, providers, and roaming conditions.
Chapters~\ref{chap:5} and~\ref{chap:6} study operator-controlled VoWiFi services using
Internet-based measurement vectors that complement radio-layer experiments.
Chapters~\ref{chap:7} and~\ref{chap:8} extend the same measurement mindset to OTT messaging applications, exposing ecosystem-wide attack surfaces and privacy risks.
Chapter~\ref{chap:4} deliberately departs from cellular measurements. Using a community-driven measurement platform, it shows how Internet centralization can be exposed and how such centralization affects anonymity in Tor.
This chapter provides an external validation of the dissertation's main theme: that accessible, large-scale measurements can uncover otherwise hidden weaknesses and risks.

\paragraph{Research questions and evidence.}
The research questions stated in Section~\ref{sec:research-questions} are addressed as follows.
RQ1 is addressed by Chapters~\ref{chap:2}--\ref{chap:3} (measurement platforms and scalability).
RQ2 is addressed by Chapters~\ref{chap:3}--\ref{chap:6} (roaming, billing, alternative access technologies).
RQ3 is addressed by Chapters~\ref{chap:5}--\ref{chap:8} (cross-layer security/privacy properties and observability),
with Chapter~\ref{chap:4} providing complementary evidence of how democratized measurements can
reveal centralization effects in a different ecosystem.

\paragraph{Artifacts and reproducibility.}
This dissertation follows an artifact-driven research approach: measurement studies are
paired with released tooling and responsible disclosure where applicable.
Section~\ref{sec:artifacts} lists the open-source projects and artifacts that support the presented experiments and facilitate independent validation, reuse, and extension by other researchers.

\paragraph{Scope.}
Although the chapters cover multiple layers and technologies, this dissertation does not aim to provide an exhaustive survey of mobile standards or deployments.
Instead, it focuses on concrete measurement capabilities and empirical case studies that collectively support the thesis statement: that independent measurements enable systematic discovery and quantification of security, privacy, and policy failures across layers.

\chapter{Relaying SIM Communication for Cellular Network Measurements}
\label{chap:2}

\begin{description}[leftmargin=4cm, style=nextline]
\item[\textbf{\large Publication Info}]

\item[\textbf{Title}] \textsc{MobileAtlas}: Geographically Decoupled Measurements in Cellular Networks for Security and Privacy Research

\item[\textbf{Authors}] \underline{Gabriel K. Gegenhuber}, Wilfried Mayer, Edgar Weippl, Adrian Dabrowski

\item[\textbf{Publication Status}] 
This paper is included in the Proceedings of the 32nd USENIX Security Symposium (USENIX Security 23), 
pp.~3493–3510, ISBN: 978-1-939133-37-3, 2023.\\
\underline{CORE2023 Ranking}: A*.

\item[\textbf{Publication Page}] \url{https://www.usenix.org/conference/usenixsecurity23/presentation/gegenhuber}

\item[\textbf{Code Artifacts}] \url{https://github.com/sbaresearch/mobile-atlas}

\item[\textbf{arXiv}] \url{https://arxiv.org/abs/2403.08507}

\item[\textbf{Reference}] \cite{gegenhuber_2023_mobileatlas}

\end{description}
\includepaper{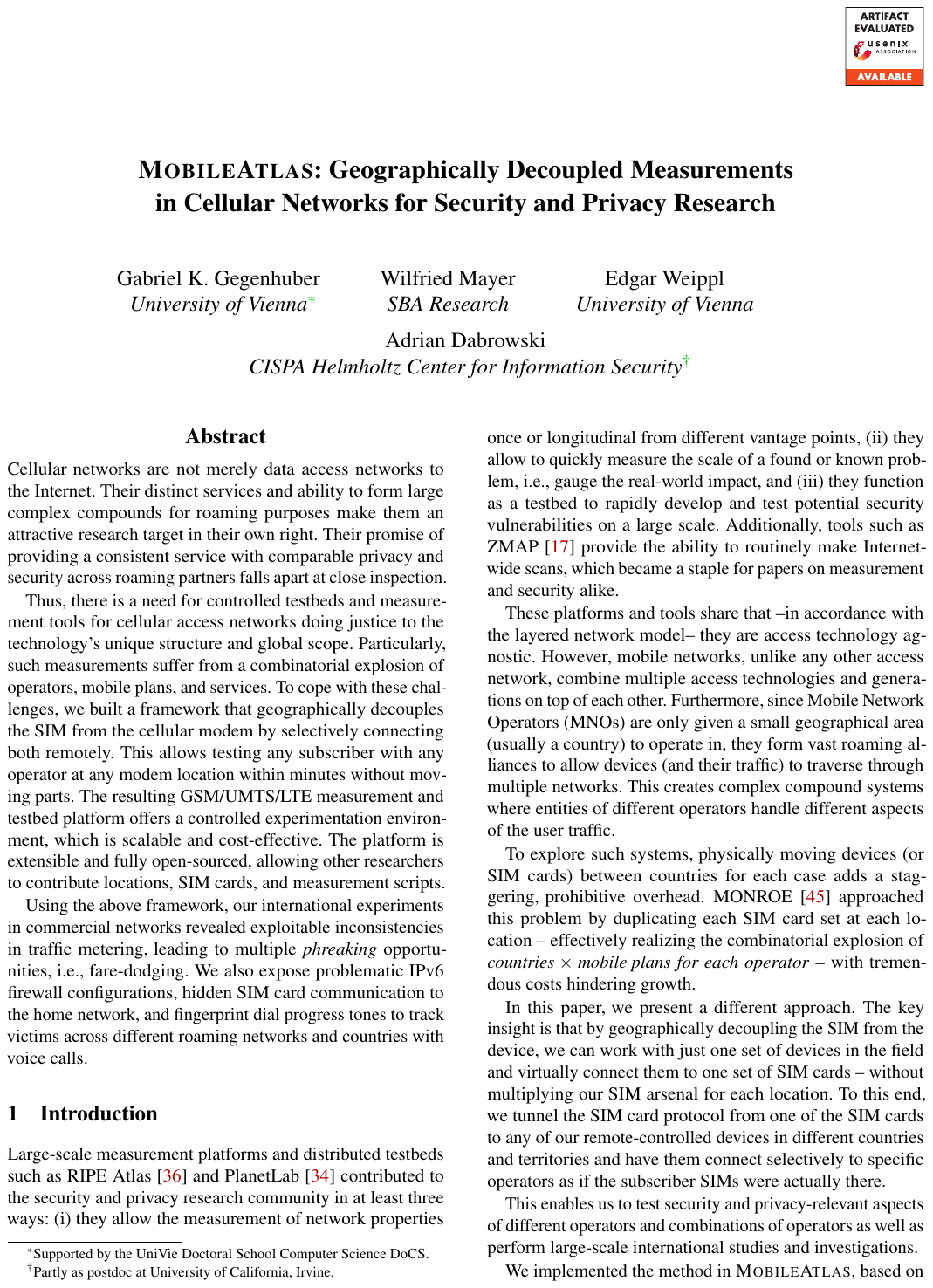}

\chapter{Measuring Traffic Classification and Zero-Rating Tariffs during International Roaming Scenarios}
\label{chap:3}

\begin{description}[leftmargin=4cm, style=nextline]
\item[\textbf{\large Publication Info}]

\item[\textbf{Title}] Zero-Rating, One Big Mess: Analyzing Differential Pricing Practices of European MNOs

\item[\textbf{Authors}] \underline{Gabriel K. Gegenhuber}, Wilfried Mayer, Edgar Weippl

\item[\textbf{Publication Status}] 
This paper is included in the Proceedings of the IEEE Global Communications Conference (GLOBECOM 2022), pp.~203--208, 2022.\\
\underline{CORE2023 Ranking}: B.

\item[\textbf{DOI}] \url{https://doi.org/10.1109/GLOBECOM48099.2022.10001701}

\item[\textbf{Code Artifacts}] \url{https://github.com/sbaresearch/mobile-atlas}

\item[\textbf{arXiv}] \url{https://arxiv.org/abs/2403.08066}

\item[\textbf{Reference}] \cite{gegenhuber_2022_zero}
\end{description}
\includepaper{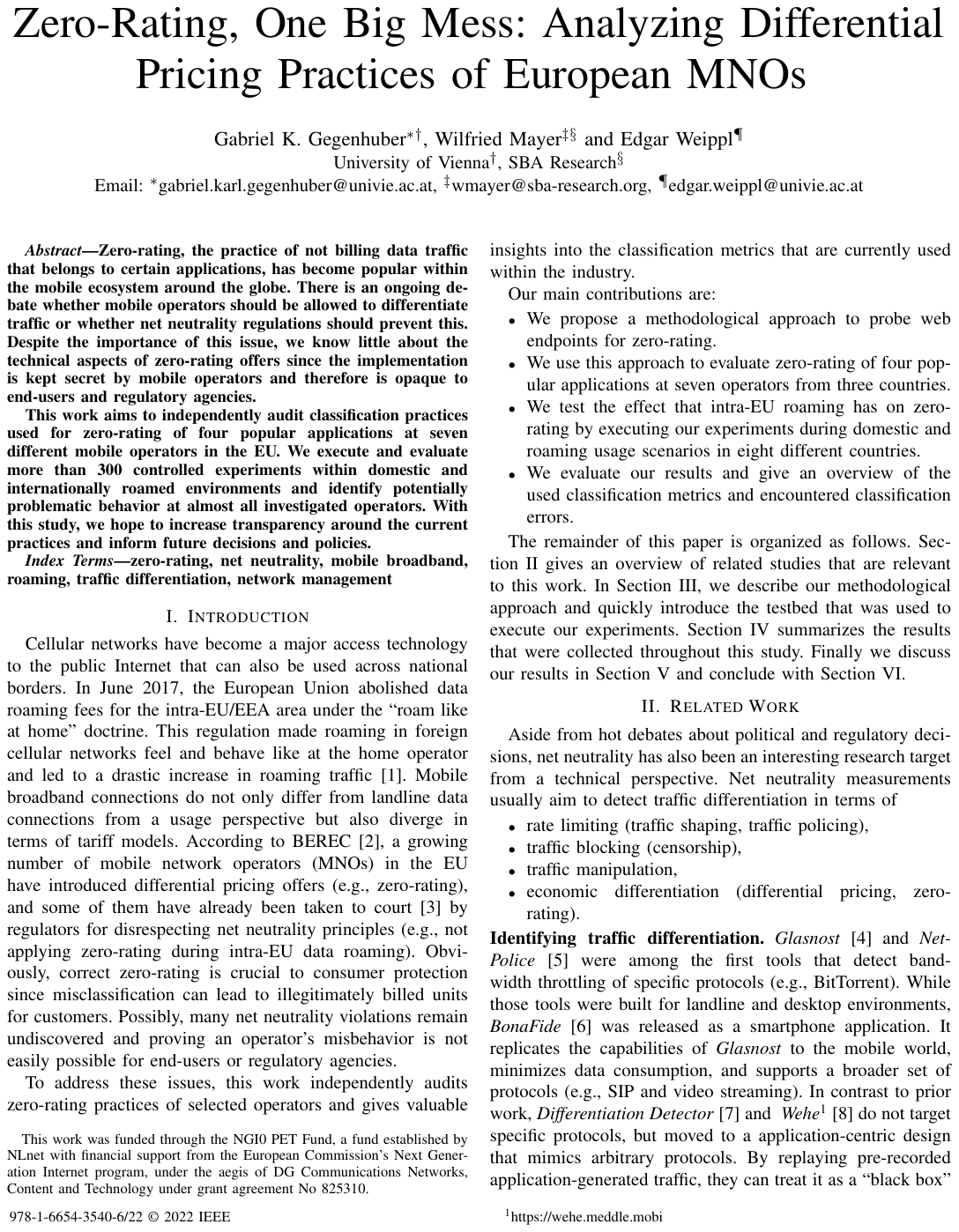}

\chapter{Detecting AS-level Centralization in Tor Using Democratized Traceroute Measurements}
\label{chap:4}

\begin{description}[leftmargin=4cm, style=nextline]
\item[\textbf{\large Publication Info}]

\item[\textbf{Title}] An Extended View on Measuring Tor AS-level Adversaries

\item[\textbf{Authors}] \underline{Gabriel K. Gegenhuber}, Markus Maier, Florian Holzbauer, Wilfried Mayer, Georg Merzdovnik, Edgar Weippl, Johanna Ullrich.

\item[\textbf{Publication Status}] 
This paper is included in Computers \& Security, Volume 132, Article 103302, 2023.\\
\underline{SCImago Ranking}: Q1.

\item[\textbf{DOI}] \url{https://doi.org/10.1016/j.cose.2023.103302}

\item[\textbf{Code Artifacts}] \url{https://github.com/sbaresearch/ripe-tor}

\item[\textbf{arXiv}] \url{https://arxiv.org/abs/2403.08517}

\item[\textbf{Reference}] \cite{gegenhuber_2023_tor}
\end{description}
\includepdf[
pages=-,
scale=0.85,
pagecommand={},
offset=5mm 0
]{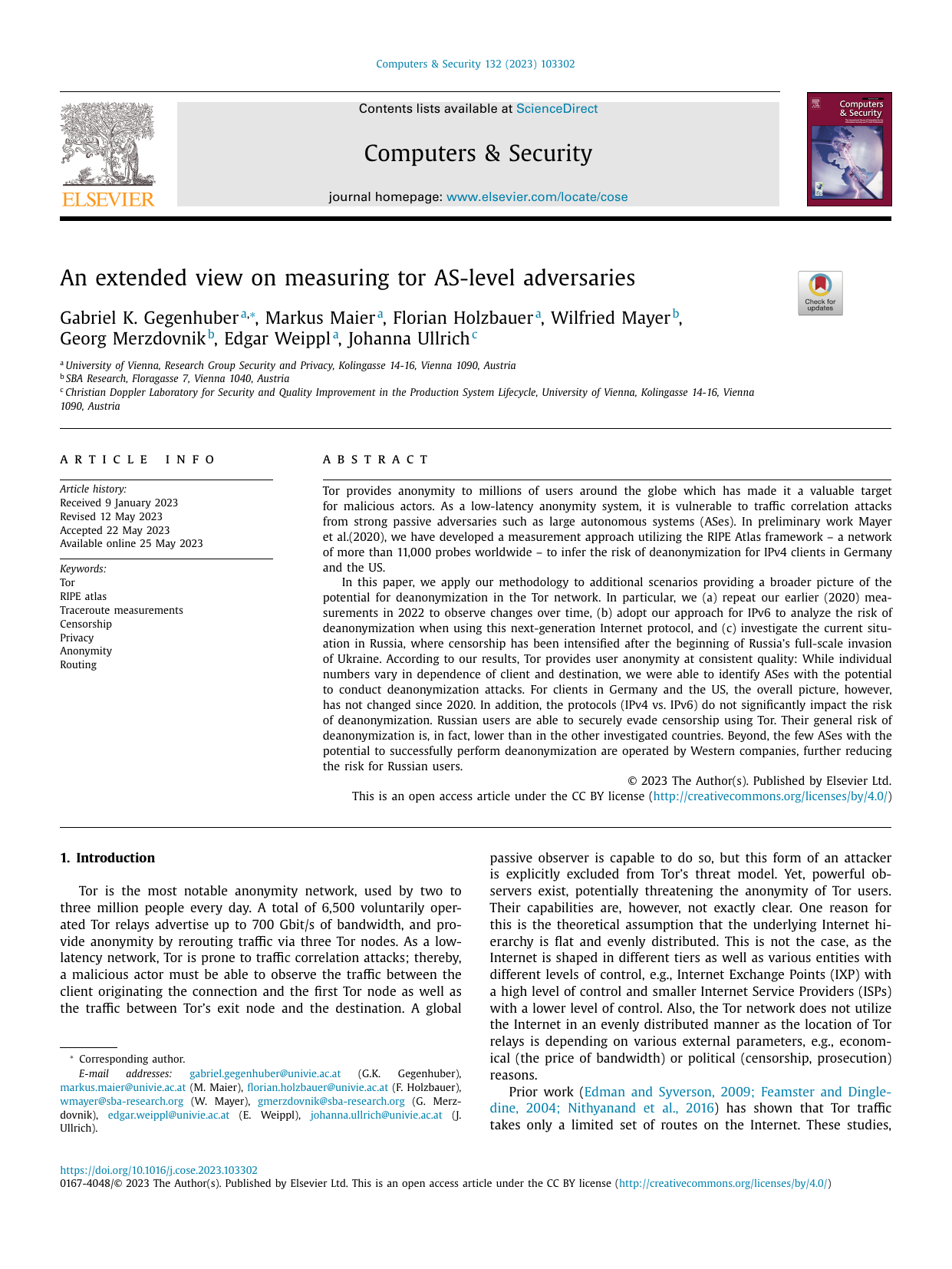}

\chapter{Measuring Geoblocking in Commercial WiFi Calling Deployments}
\label{chap:5}

\begin{description}[leftmargin=4cm, style=nextline]
\item[\textbf{\large Publication Info}]

\item[\textbf{Title}] Why E.T. Can't Phone Home: A Global View on IP-based Geoblocking at VoWiFi

\item[\textbf{Authors}] \underline{Gabriel K. Gegenhuber}, Philipp É. Frenzel, Edgar Weippl

\item[\textbf{Publication Status}] 
This paper is included in the Proceedings of the 22nd Annual International Conference on Mobile Systems, Applications and Services (MobiSys), pp.~183--195, ISBN 979-8-4007-0581-6, 2024\\
\underline{CORE2023 Ranking}: A.

\item[\textbf{DOI}] \url{https://doi.org/10.1145/3643832.3661883}

\item[\textbf{Code Artifacts}] \url{https://github.com/sbaresearch/scanywhere}

\item[\textbf{arXiv}] \url{https://arxiv.org/abs/2403.11759}

\item[\textbf{Reference}] \cite{gegenhuber_2024_geoblocking}
\end{description}
\includepaper{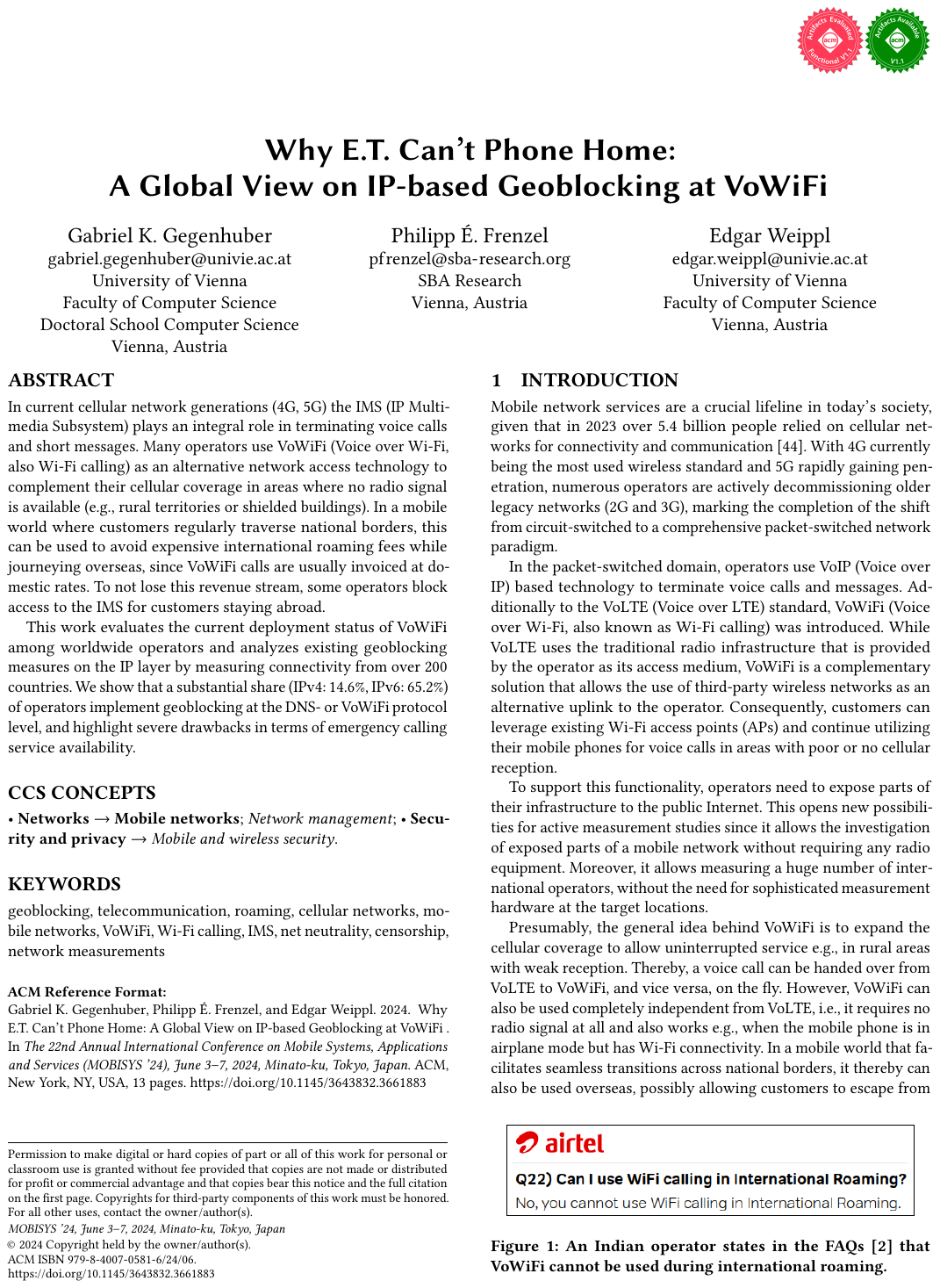}

\chapter{Measuring Insecure Configurations in Commercial WiFi Calling Deployments}
\label{chap:6}

\begin{description}[leftmargin=4cm, style=nextline]
\item[\textbf{\large Publication Info}]

\item[\textbf{Title}] Diffie-Hellman Picture Show: Key Exchange Stories from Commercial VoWiFi Deployments

\item[\textbf{Authors}] \underline{Gabriel K. Gegenhuber}, Florian Holzbauer, Philipp Frenzel, Edgar Weippl, Adrian Dabrowski

\item[\textbf{Publication Status}] 
This paper is included in the Proceedings of the 33rd USENIX Security Symposium (USENIX Security 24), 
pp.~451--468, ISBN 978-1-939133-44-1, 2024.\\
\underline{CORE2023 Ranking}: A*.

\item[\textbf{Publication Page}] \url{https://www.usenix.org/conference/usenixsecurity24/presentation/gegenhuber}

\item[\textbf{Code Artifacts}] \url{https://github.com/sbaresearch/vowifi-epdg-scanning}\\
\url{https://github.com/sbaresearch/mbn-mcfg-tools}

\item[\textbf{arXiv}] \url{https://arxiv.org/abs/2407.19556}

\item[\textbf{Reference}] \cite{gegenhuber_2024_diffie}
\end{description}
\includepaper{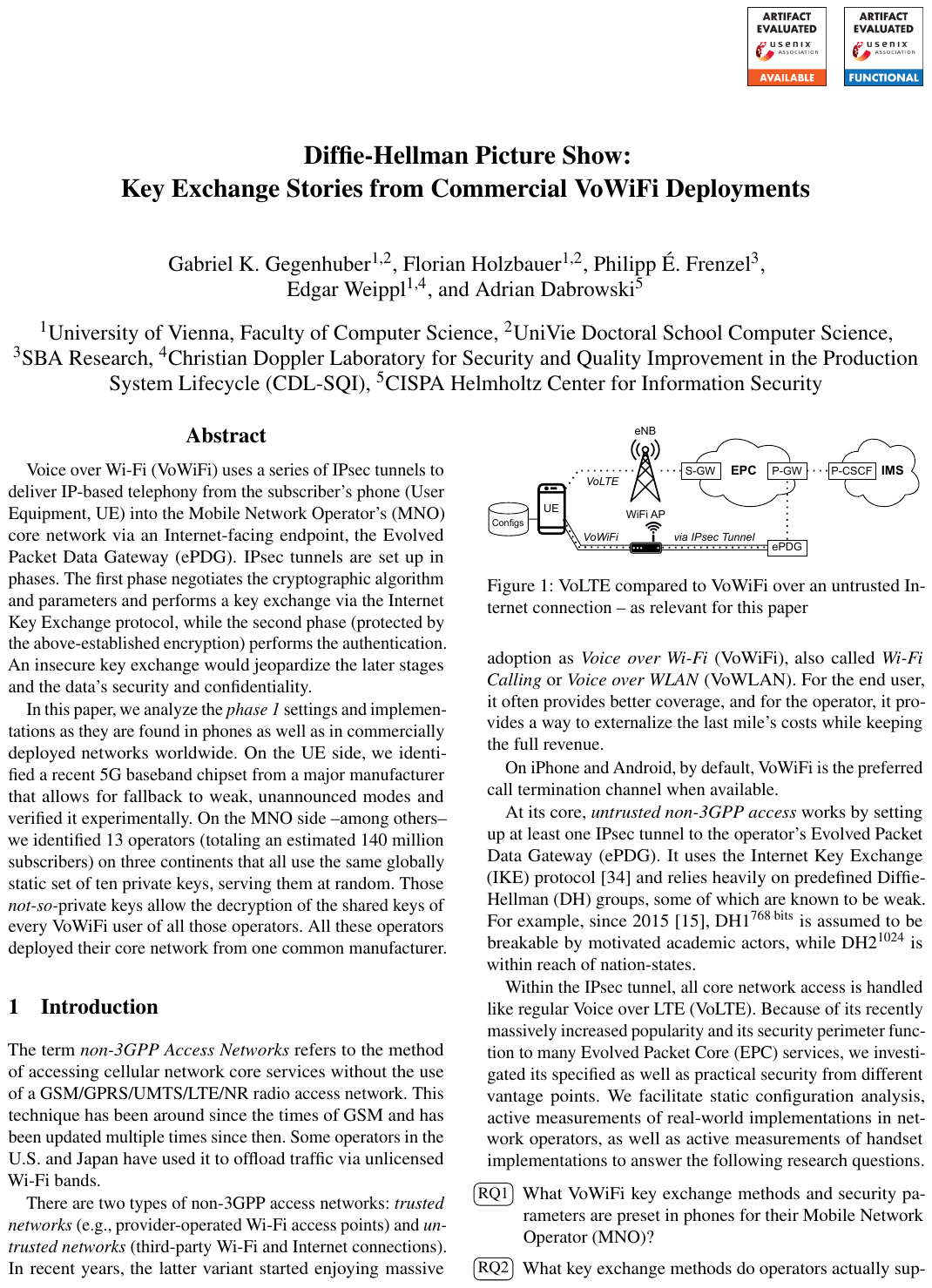}

\chapter{Individual User Monitoring via Silent Pings on Instant Messengers}
\label{chap:7}

\begin{description}[leftmargin=4cm, style=nextline]
\item[\textbf{\large Publication Info}]

\item[\textbf{Title}] Careless Whisper: Exploiting Silent Delivery Receipts to Monitor Users on Mobile Instant Messengers

\item[\textbf{Authors}] \underline{Gabriel K. Gegenhuber}, Maximilian Günther, Markus Maier, Aljosha Judmayer, Florian Holzbauer, Philipp É. Frenzel, Johanna Ullrich

\item[\textbf{Publication Status}] 
This paper is included in the Proceedings of the 28th International Symposium on Research in Attacks, Intrusions and Defenses (RAID), 2025. 
To appear in the proceedings.
\textit{Distinguished with the Best Paper Award.}\\
\underline{CORE2023 Ranking}: A.

\item[\textbf{DOI}] To be assigned by the publisher.

\item[\textbf{arXiv}] \url{https://arxiv.org/abs/2411.11194}

\item[\textbf{Reference}] \cite{gegenhuber_2025_carelesswhisper}
\end{description}
\includepaper{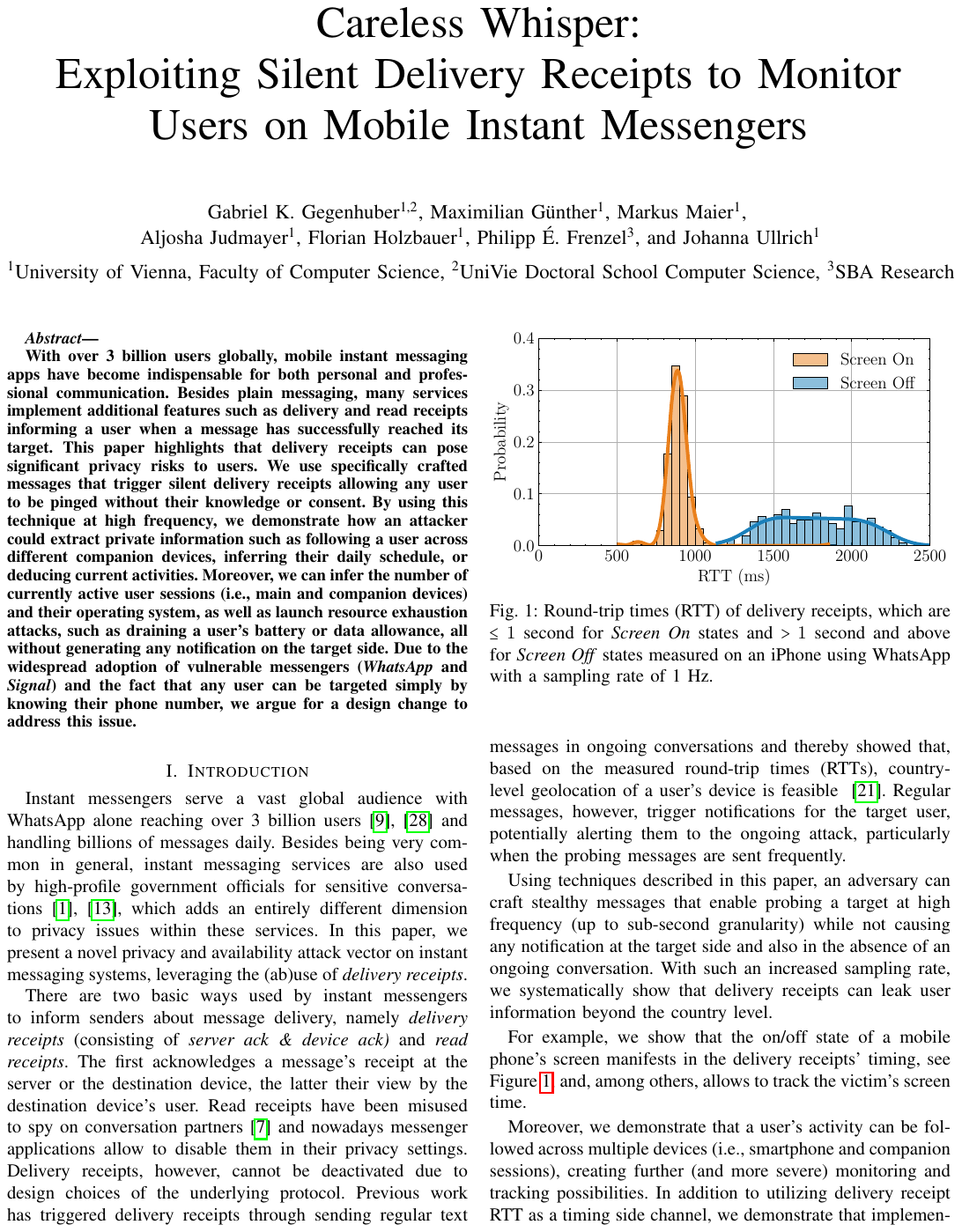}

\chapter{Exploits via E2EE Prekeying Mechanism on Instant Messengers}
\label{chap:8}

\begin{description}[leftmargin=4cm, style=nextline]
\item[\textbf{\large Publication Info}]

\item[\textbf{Title}] Prekey Pogo: Investigating Security and Privacy Issues in WhatsApp's Handshake Mechanism

\item[\textbf{Authors}] \underline{Gabriel K. Gegenhuber}, Philipp É. Frenzel, Maximilian Günther, Aljosha Judmayer

\item[\textbf{Publication Status}] 
This paper is included in the Proceedings of the Proceedings of the 19th USENIX Conference on Offensive Technologies (WOOT 25), 
pp.~209--227, 2025.\\

\item[\textbf{Publication Page}] \url{https://www.usenix.org/conference/woot25/presentation/gegenhuber}

\item[\textbf{Code Artifacts}] \url{https://github.com/sbaresearch/prekey-pogo}

\item[\textbf{arXiv}] \url{https://arxiv.org/abs/2504.07323}

\item[\textbf{Reference}] \cite{gegenhuber_2025_prekeypogo}
\end{description}
\includepaper{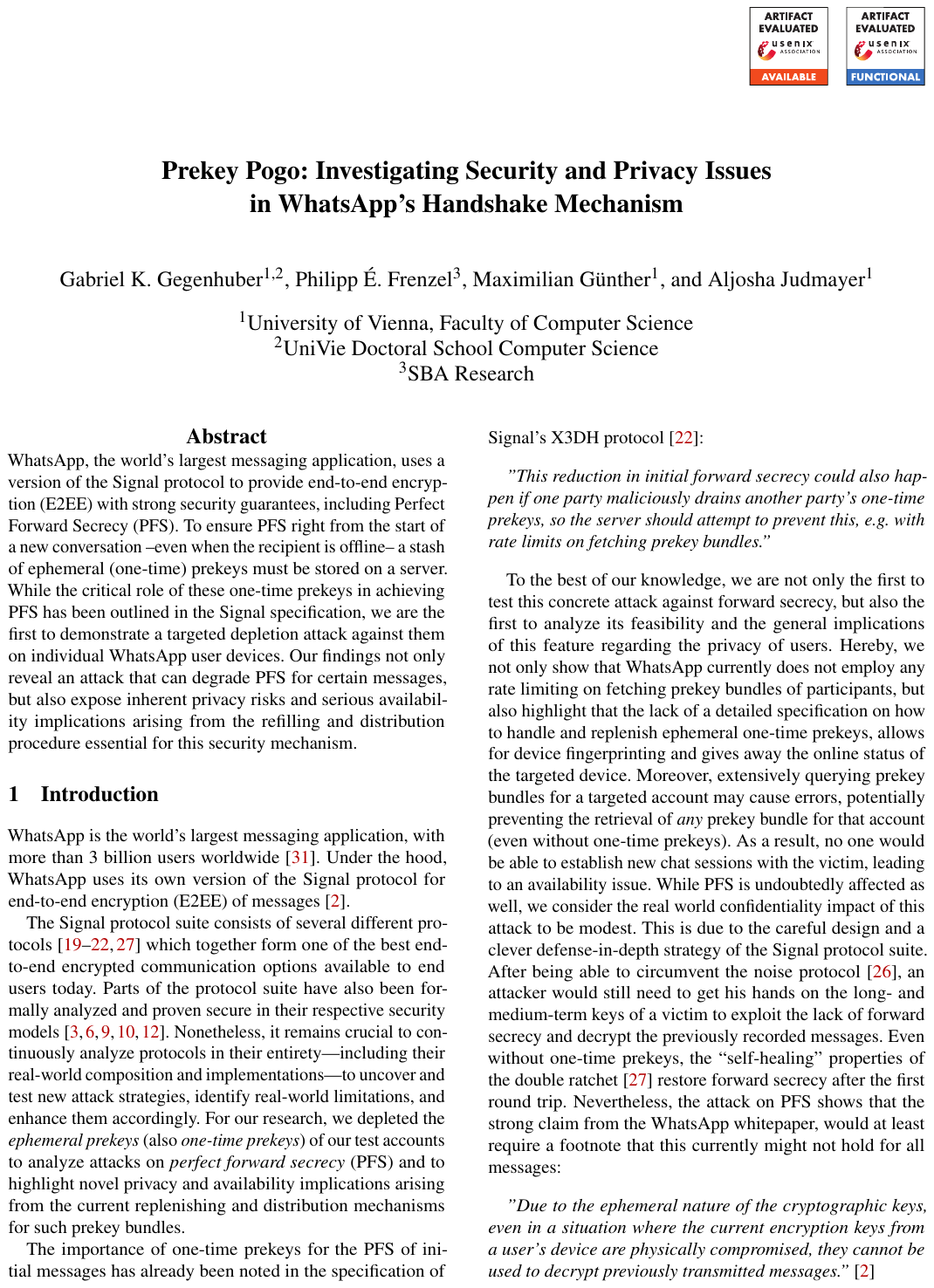}

    \chapter{Conclusion}
\label{chap:9} 

Cellular networks form a resilient global communication infrastructure due to their inherently distributed architecture.
Radio access networks and VoWiFi services are operated by independent entities across geographic and administrative boundaries, reducing the impact of failures to a limited region and user base.

At the same time, this dissertation shows that many components of the mobile communication ecosystem have become increasingly vulnerable and centralized.
Internet-accessible operator gateways expand the attack surface and are frequently operated with outdated or insecure configurations.
Moreover, globally dominant OTT instant messaging platforms consolidate functionality that was previously distributed, introducing systemic risks by concentrating failure domains and attack surfaces at a massive scale.
This risk is not only theoretical: recent work reports that roughly 3.5 billion WhatsApp accounts could be enumerated, illustrating how a single service can create ecosystem-wide exposure~\cite{gegenhuber_2025_heythere}.

Across all settings, this dissertation demonstrates that active, independent measurements are essential to uncover and address such risks.
By introducing \textsc{MobileAtlas} and leveraging Internet-based measurement vectors, this dissertation enables controlled and reproducible studies across cellular access networks, operator-managed services, and third-party communication platforms.
The presented measurements show that while decentralization enhances resilience, centralization simplifies observation, control, and, consequently, exploitation.

Beyond individual vulnerabilities, this work challenges the assumption that standardization alone provides security and privacy guarantees in mobile communication systems.
The presented measurements demonstrate that without independent verification, such assumptions can mask systemic weaknesses across operators and platforms.

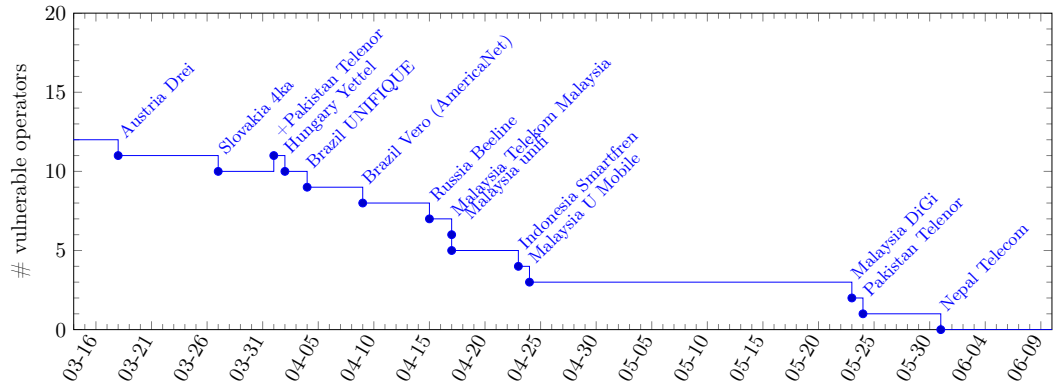
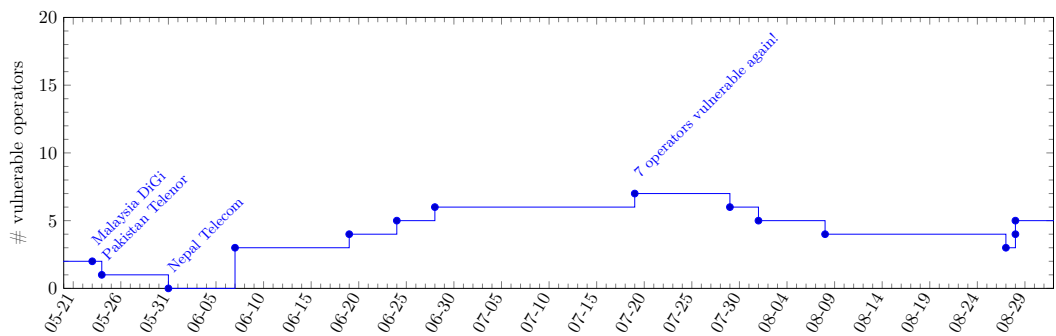
\begin{figure*}[!t]
    \centering

    \begin{subfigure}{0.99\linewidth}
        \centering
        \resizebox{\linewidth}{!}{%
            \begin{tikzpicture}%

  \begin{axis}[date coordinates in=x,date ZERO=2024-03-01,
               xticklabel=\month-\day,
               xticklabel style={rotate=60,anchor=east},
               minor x tick num=4,
               minor y tick num=4,
               ymin=0,ymax=20,
       xmin=2024-03-14,
       xmax=2024-06-10, 
       ylabel={\# vulnerable operators},
       x=2.0mm]

\addplot+[const plot mark left,
    every node near coord/.append style={xshift=-1pt,yshift=8pt,anchor=west,font=\footnotesize,rotate=45},
    nodes near coords={\labelz}, 
    visualization depends on={value \thisrowno{2}\as\labelz}] 
table[col sep=comma,x=date,y=num]
{
date,    num, name,    mccmnc
2024-01-01,12,,
2024-03-18,11,	Austria Drei,	232-05 232-10
2024-03-27,10,	Slovakia 4ka,	231-03
2024-04-01,11,	+Pakistan Telenor,	410-06
2024-04-02,10,	Hungary Yettel,	216-01
2024-04-04,9,	Brazil UNIFIQUE,	724-29
2024-04-09,8,	Brazil Vero (AmericaNet),	724-26
2024-04-15,7,	Russia Beeline,	250-99
2024-04-17,6,	Malaysia Telekom Malaysia,	502-11
2024-04-17,5,	~~~Malaysia unifi,	502-153
2024-04-23,4,	Indonesia Smartfren	,510-09 510-28
2024-04-24,3,	Malaysia U Mobile,	502-18 
2024-05-23,2,	Malaysia DiGi,	502-16
2024-05-24,1,	Pakistan Telenor,	410-06
2024-05-31,0,   Nepal Telecom, 429-01
2024-12-31,0,,
};
 
\end{axis}
\end{tikzpicture}%
        }
        \caption{\textbf{Initial remediation timeline}. After coordinated disclosure via the GSMA, vulnerable operators gradually replaced the reused private-keys used for the IKE/IPSec handshake over a period of several months, eventually eliminating the issue across all observed deployments.}
        \label{fig:remediation:a}
    \end{subfigure}

    \vspace{2ex}

    \begin{subfigure}{0.99\linewidth}
        \centering
        \resizebox{\linewidth}{!}{%
            \begin{tikzpicture}%

  \begin{axis}[date coordinates in=x,date ZERO=2024-05-01,
               xticklabel=\month-\day,
               xticklabel style={rotate=60,anchor=east},
               minor x tick num=4,
               minor y tick num=4,
               ymin=0,ymax=20,
       xmin=2024-05-20,
       xmax=2024-09-01, 
       ylabel={\# vulnerable operators},
       x=2.0mm]

\addplot+[const plot mark left,
    every node near coord/.append style={xshift=-1pt,yshift=8pt,anchor=west,font=\footnotesize,rotate=45},
    nodes near coords={\labelz}, 
    visualization depends on={value \thisrowno{2}\as\labelz}] 
table[col sep=comma,x=date,y=num]
{
date,    num, name,    mccmnc
2024-01-01,2,,
2024-05-23,2,	Malaysia DiGi,	502-16
2024-05-24,1,	Pakistan Telenor,	410-06
2024-05-31,0,   Nepal Telecom, 429-01
2024-06-05 2, 
2024-06-07,3,  
2024-06-19,4, 
2024-06-24,5, 
2024-06-28,6
2024-07-19,7, 7 operators vulnerable again!
2024-07-29,6,
2024-08-01,5,
2024-08-08,4,
2024-08-27,3,
2024-08-28,4,
2024-08-28,5,
2024-09-12,4,
2024-12-31,4,,
};
 
\end{axis}
\end{tikzpicture}%
        }
        \caption{\textbf{Subsequent regression timeline}. Follow-up ePDG scans of global operators revealed a reappearance of reused private-keys during the IKE/IPsec handshake shortly after full remediation, affecting both previously remediated operators and newly observed vulnerable deployments.}
        \label{fig:remediation:b}
    \end{subfigure}

    \vspace{1ex}
    \caption{Global reuse of static private keys in ePDG deployments: remediation and regression over time. Without continuous, worldwide scans of operator infrastructures, neither the original vulnerability nor its subsequent reoccurrence would have been discovered.}
    \label{fig:remediation}
\end{figure*}

An interesting trade-off that was observed throughout this work concerns mitigation and patch deployment.
Centralized services set clear responsibilities, allow for rapid patching and coordinated security updates, often mitigating vulnerabilities within several days.
In contrast, decentralized cellular infrastructures require patches and configuration changes to be rolled out across many independently operated networks, resulting in slower and more uneven adoption, as we observed for \textit{CVE-2024-22064} (shown in Figure~\ref{fig:remediation}).
While this longer rollout slope can delay mitigation, it also limits the blast radius of individual failures.

\smallskip
\paragraph{Future directions.}
This dissertation does not aim to provide an exhaustive survey of mobile standards or deployments.
Consequently, future measurement work should also cover the protocol and service alternatives that increasingly carry everyday communication:
Apple-controlled iMessage has become a de facto default messaging channel on iOS;
Rich Communication Services (RCS), originally developed as an open standard but nowadays heavily influenced (and, in practice, often controlled) by Google (e.g., because many operators rely on Google-hosted backends via Jibe);
and the email ecosystem, which has consolidated around a few large providers (e.g., Gmail, Outlook) while smaller providers continue to diminish and often lag in deploying available security mechanisms (e.g., DNSSEC)~\cite{holzbauer_2022_notsimple}.
Beyond technical risk, the concentration of these backends outside the EU raises digital sovereignty concerns by creating dependencies on external providers and jurisdictions.
Extending the dissertation's methodology to map backend centralization, configuration hygiene, and metadata leakage across these systems, and to track mitigation rollouts longitudinally, would strengthen our ability to detect emerging systemic risks early.

\bigskip
In summary, this dissertation highlights a fundamental tension in modern mobile communication systems: decentralization strengthens resilience, while centralization amplifies both risk and control.
Independent, scalable measurement capabilities are indispensable to understanding this balance, detecting emerging systemic vulnerabilities, and supporting informed decisions that improve the security, privacy, and robustness of global communication infrastructures.

    \bibliographystyle{alpha}
    \bibliography{bibliography}

\newcommand{\etalchar}[1]{$^{#1}$}
\begin{thebibliography}{GMWD23}

\bibitem[AWR14]{anderson_2014_global}
Collin Anderson, Philipp Winter, and Roya.
\newblock {Global Network Interference Detection Over the RIPE Atlas Network}.
\newblock In {\em {Workshop on Free and Open Communications on the Internet (FOCI)}}, 2014.

\bibitem[blu03]{bluetooth_sim_access_2003}
{Bluetooth SIM Access Profile Specification}.
\newblock Technical report, 3GPP, 2003.

\bibitem[Geg25]{gegenhuber_2025_novel}
Gabriel~K. Gegenhuber.
\newblock {Security and Privacy Measurements in Cellular Networks: Novel Approaches in a Global Roaming Context}.
\newblock In {\em 32nd ACM Conference on Computer and Communications Security (CCS)}, 2025.

\bibitem[GF25]{gegenhuber_2025_scanywhere}
Gabriel~K. Gegenhuber and Philipp~{\'E}. Frenzel.
\newblock {Scanywhere: Distributed Internet Scanning Leveraging Commercial {VPN} Subscriptions}.
\newblock In {\em 9th Network Traffic Measurement and Analysis Conference (TMA)}, 2025.

\bibitem[GFD25]{gegenhuber_2025_simulator}
Gabriel~K. Gegenhuber, Philipp~{\'E}. Frenzel, and Adrian Dabrowski.
\newblock {SIMulator: {SIM} Tracing on a (Pico-)Budget}.
\newblock In {\em 18th ACM Conference on Security and Privacy in Wireless and Mobile Networks (WiSec)}, 2025.

\bibitem[GFG{\etalchar{+}}26]{gegenhuber_2025_heythere}
Gabriel~K. Gegenhuber, Philipp~{\'E}. Frenzel, Maximilian G{\"u}nther, Johanna Ullrich, and Aljosha Judmayer.
\newblock {Hey there! You are using WhatsApp: Enumerating Three Billion Accounts for Security and Privacy}.
\newblock In {\em 33rd Annual Network and Distributed System Security Symposium (NDSS)}, 2026.

\bibitem[GFGJ25]{gegenhuber_2025_prekeypogo}
Gabriel~K. Gegenhuber, Philipp~{\'E}. Frenzel, Maximilian G{\"u}nther, and Aljosha Judmayer.
\newblock {Prekey Pogo: Investigating Security and Privacy Issues in WhatsApp's Handshake Mechanism}.
\newblock In {\em 19th USENIX WOOT Conference on Offensive Technologies (WOOT)}, 2025.

\bibitem[GFW24a]{gegenhuber_2024_never}
Gabriel~K. Gegenhuber, Philipp~{\'E}. Frenzel, and Edgar Weippl.
\newblock {Never Gonna Give You Up: Exploring Deprecated {NULL} Ciphers in Commercial {VoWiFi} Deployments}.
\newblock In {\em 17th ACM Conference on Security and Privacy in Wireless and Mobile Networks (WiSec)}, 2024.

\bibitem[GFW24b]{gegenhuber_2024_geoblocking}
Gabriel~K. Gegenhuber, Philipp~É. Frenzel, and Edgar Weippl.
\newblock {Why E.T. Can't Phone Home: A Global View on IP-based Geoblocking at VoWiFi}.
\newblock In {\em Proceedings of the 22nd Annual International Conference on Mobile Systems, Applications, and Services (MobiSys 2024)}, 2024.

\bibitem[GGM{\etalchar{+}}25]{gegenhuber_2025_carelesswhisper}
Gabriel~K. Gegenhuber, Maximilian G{\"u}nther, Markus Maier, Aljosha Judmayer, Florian Holzbauer, Philipp~{\'E}. Frenzel, and Johanna Ullrich.
\newblock {Careless Whisper: Exploiting Silent Delivery Receipts to Monitor Users on Mobile Instant Messengers}.
\newblock In {\em 28th International Symposium on Research in Attacks, Intrusions and Defenses (RAID)}, 2025.

\bibitem[GHF{\etalchar{+}}24]{gegenhuber_2024_diffie}
Gabriel~K. Gegenhuber, Florian Holzbauer, Philipp~{\'E}. Frenzel, Edgar Weippl, and Adrian Dabrowski.
\newblock {{Diffie-Hellman} Picture Show: Key Exchange Stories from Commercial {VoWiFi} Deployments}.
\newblock In {\em 33rd USENIX Security Symposium (USENIX Security)}, 2024.

\bibitem[GLHS25]{gegenhuber_2025_airtag}
Gabriel~K. Gegenhuber, Leonid Liadveikin, Florian Holzbauer, and Sebastian Strobl.
\newblock {A Relay a Day Keeps the {AirTag} Away: Practical Relay Attacks on Apple's {AirTags}}.
\newblock In {\em 41st Annual Computer Security Applications Conference (ACSAC)}, 2025.

\bibitem[GMH{\etalchar{+}}23]{gegenhuber_2023_tor}
Gabriel~K. Gegenhuber, Markus Maier, Florian Holzbauer, Wilfried Mayer, Georg Merzdovnik, Edgar Weippl, and Johanna Ullrich.
\newblock {An extended view on measuring tor as-level adversaries}.
\newblock {\em Computers \& Security}, 132:103302, 2023.

\bibitem[GMW22]{gegenhuber_2022_zero}
Gabriel~K. Gegenhuber, Wilfried Mayer, and Edgar Weippl.
\newblock {Zero-Rating, One Big Mess: Analyzing Differential Pricing Practices of European MNOs}.
\newblock In {\em IEEE Global Communications Conference (GLOBECOM)}, 2022.

\bibitem[GMWD23]{gegenhuber_2023_mobileatlas}
Gabriel~K. Gegenhuber, Wilfried Mayer, Edgar Weippl, and Adrian Dabrowski.
\newblock {MobileAtlas: Geographically Decoupled Measurements in Cellular Networks for Security and Privacy Research}.
\newblock In {\em 32nd USENIX Security Symposium (USENIX Security)}, 2023.

\bibitem[HULF22]{holzbauer_2022_notsimple}
Florian Holzbauer, Johanna Ullrich, Martina Lindorfer, and Tobias Fiebig.
\newblock {Not that Simple: Email Delivery in the 21st Century}.
\newblock In {\em USENIX Annual Technical Conference (USENIX ATC)}, 2022.

\bibitem[ISO06]{iso7816-3}
{ISO/IEC 7816-3:2006 - Identification cards — Integrated circuit cards — Part 3: Cards with contacts — Electrical interface and transmission protocols}.
\newblock Standard, International Organization for Standardization, Geneva, CH, November 2006.

\bibitem[{RIP}15]{staff_2015_ripe}
{RIPE NCC Staff}.
\newblock {RIPE Atlas: A Global Internet Measurement Network}.
\newblock {\em Internet Protocol Journal}, 18(3), 2015.

\bibitem[SB24]{gsma_2024_state}
Matthew Shanahan and Kalvin Bahia.
\newblock {The State of Mobile Internet Connectivity 2024}.
\newblock Technical report, GSM Association, 2024.

\end{thebibliography}

    \cleardoublepage{}
	\pagebreak

\end{document}